\documentclass{emulateapj}
\usepackage{amssymb,amsmath}
\usepackage{graphicx}
\usepackage{natbib}
\begin{document}
\title{Balancing the load: A Voronoi based scheme for parallel computations}
\author{Elad Steinberg\altaffilmark{1}, Almog Yalinewich\altaffilmark{1}, Re'em Sari\altaffilmark{1} and Paul Duffell\altaffilmark{2}}
\altaffiltext{1}{Racah Institute of Physics, Hebrew University, Jerusalem 91904, Israel}
\altaffiltext{2}{Center for Cosmology and Particle Physics, New York University, New York, NY, USA}

\email{elad.steinberg@mail.huji.ac.il}
%\address[heb]{Racah Institute of Physics, Hebrew University, Jerusalem 91904, %Israel}
%\address[nyu]{Center for Cosmology and Particle Physics, New York University, New %York, NY, USA}
\begin{abstract}
One of the key issues when running a simulation on multiple CPUs is maintaining a proper load balance throughout the run and minimizing communications between CPUs.
We propose a novel method of utilizing a Voronoi diagram to achieve a  nearly perfect load balance without the need of any global redistributions of data. As a show case, we implement our method in RICH, a 2D moving mesh hydrodynamical code, but it can be extended trivially to other codes in 2D or 3D. Our tests show that this method is indeed efficient and can be used in a large variety of existing hydrodynamical codes.
\end{abstract}

\section{Introduction}
Large computational tasks require machines with distributed memory, since in such systems, the number of CPUs and the size of the total available memory is not set by the capacity of a single machine.
An important downside to the distributed memory approach is that communications between processors may take a significant fraction of the computation time since information is carried over a network.

Ensuring proper load balance during parallel calculations of partial differential equations (e.g. hydrodynamics) using the distributed memory approach, is essential for achieving superior performance. In this paper we present a general scheme that is applicable to any problem that can be solved with domain decomposition. Since our implementation is for hydrodynamics, the rest of the paper uses nomenclature of hydrodynamics without loss of generality.
%Distributing the workload and data for SPH or non uniform grid based codes is non-trivial and a time consuming task.
For uniform fixed mesh codes (e.g. Athena \citep{Athena}) load balancing is no great challenge. Partitioning the domain into equal sized static partitions is straightforward, results in balanced loading and minimal communications.
For static but nonuniform mesh, a tree based partitioning can be used (e.g. \cite{Dubinski}) to divide the domain among the different CPUs.
Smooth particle hydrodynamics (SPH), Adaptive Mesh Refinement (AMR) and moving mesh codes differ from the above since the number of cells/points in a given part of the domain is not constant in time.
Pluto \citep{Pluto}, an AMR code, achieves load balancing by implementing the Kernighan-Lin algorithm for solving knapsack problems (see \cite{book} for more details).
Recently it has become very popular to use space filling curves (e.g. Morton curve \citep{morton} or a Peano-Hilbert curve \citep{sb}) to map the cells/points among the different CPUs and to define their boundaries \citep{sb,Gadget,Ramses}.
However, as the points move during the time evolution, the load balance between processors deteriorates and redistribution among CPUs is necessary.

\section{A Voronoi based parallelization scheme}
The novel method introduced here uses a Voronoi diagram to decompose the domain. This is not to be confused with hydrodynamic Voronoi schemes, like AREPO or TESS \citep{Arepo,Tess}, or own code RICH \citep{Rich}, which use Voronoi diagram to determine the computational cells boundaries.
The basic building blocks of the method are CPU mesh points. The corresponding Voronoi diagram based on these mesh points defines for each CPU a region within the computational domain.  Each CPU retains in its memory only the hydro points that lie inside its Voronoi cell. At the beginning of the calculation, the domain is decomposed via the Voronoi diagram, and each CPU is assigned its relevant hydro points. The Voronoi diagram is built only for the CPU mesh points, while the hydro points (or cells for grid based codes) have their own independent structure that is the same as it would be in the serial version.

At the beginning of each subsequent time step, we check which hydro points moved between different CPU Voronoi cells, and send their information to the new CPU in which they lie now.
This can be done extremely easily and efficiently since the Voronoi diagram already contains the information of who are the neighbors of each CPU.
Since typically hydro points can not move in a single time step distances that are larger than their size, the number of communications required is limited to about $\mathcal{O}(\sqrt{N})$ for 2D or $\mathcal{O}(N^{2/3})$ for 3D,
%between neighboring CPUs
where $N$ is the number of hydro points per CPU. Therefore, for large enough $N$ the communication time is guaranteed to be a small fraction of the computational time even if communications are relatively slow.

%This scheme is great for the first time step, where near-perfect load balancing can be %achieved by having the distribution of the CPU mesh points mimic the distribution of the %hydro points. However, after a few time steps the hydro points tend to move around and %worsen the load balance.

Another novelty of this method is that the CPU mesh points can be moved in such a manner as to try and maintain a constant workload per CPU.
To determine how the CPU mesh points are to be moved, each CPU is given a merit based on its current workload.
%This scheme has no need for global redistribution of data among CPUs in order to maintain a good load balance, which is very time consuming, rather only local exchanges between neighboring CPUs

Specifically, we following a scheme which we call ``Pressure Balancing Scheme''. For each time step (or every few time steps if the problem is ``smooth" with regards to its hydro points distribution) we implement the following procedure:
\begin{enumerate}
  \item Every CPU constructs the Voronoi diagram of all of the CPU mesh points.
  \item The information of Hydro points that moved outside the CPU's Voronoi cell to which they belonged in the previous time step are sent to their new CPU and new hydro points are received from the neighboring CPUs.
  \item A hydrodynamical time step is preformed, during which ghost hydro points are communicated to/from neighboring CPUs. Since in a 2D Voronoi diagram a point has on average 6 neighbors, each CPU in our implementation communicates on average with 12 other CPUs (its adjacent neighbors and the neighbors mutual neighbors). For codes that hold a tree data structure for the location of the points or grid based codes, this number can be reduced in most cases only to the adjacent neighbors. In the case of Voronoi based moving mesh codes, a Voronoi diagram of all of the local hydro points is constructed taking into account information from the received ghost hydro points. This is preformed in a separate communication than the one in the above bullet.
  \item At the end of the computational time step, each CPU determines its merit, which in our scheme we call ``pressure", by counting how many hydro points are in its domain and this data is broadcasted to the neighboring CPUs.
  \item Every CPU determines its new location by
  \begin{equation}
  \begin{split}
  \vec{dx_i}&=M_{\mathrm{best}}\cdot\sum_{j=0}^{\mathrm{Neighbors}}(\vec{x}_i^n-\vec{x}_j^n)(1/(M_j+1)\\
  &-1/(M_i+1))
  \end{split}
  \end{equation}
  \begin{equation}
  \vec{x}_i^{n+1}=\vec{x}_i^n+\hat{dx_i}\cdot\mathrm{Min}(\alpha R_i,\mathrm{abs}(\vec{dx_i}))
  \end{equation}
  where $M_{\mathrm{best}}$ is the total number of hydro points divided by the number of processors, $M_i$ is the number of hydro points in the $\mathrm{i}-th$ CPU. $R_i$ is the CPU's effective radius calculated as $\pi R_i^2=A_i$ for 2D and $4\pi R_i^3/3=V_i$ for 3D where $A_i$ is the area of the $\mathrm{i}-th$ CPU Voronoi cell in 2D and $V_i$ is its volume in 3D, $n$ is the temporal index and we set $\alpha=0.04$. The worst case scenario for the fractional number of hydro points that are to be communicated in the next time step is of order $\alpha$.
  \item Optionally, a Lloyd iteration can be performed (e.g. eq 63 in \cite{Arepo}) in order to ensure that the CPU cells remain rather round.
\end{enumerate}
This scheme allows the domains of CPUs to ``drift'' with the flow of the points, maintaining a good load balance and diminishing the amount of data exchange between CPUs.

In order to achieve best performance, the initial positions of the CPU mesh points should mimic the distribution of the hydro points. If the initial load balance is poor, or an abrupt shift in the load balance occurs, the number of time steps to regain good load balance can be roughly estimated as the number of time steps it takes a CPU mesh point to travel the distance of the domain size. If $A$ is the area of the domain, then the distance that a badly load balanced CPU mesh point has to travel, $\Delta r$, is of order
\begin{equation}
\Delta r\approx \sqrt{A}.
\end{equation}
The distance that a CPU mesh point travels in a single time step, $\delta r$, is roughly
\begin{equation}
\delta r\approx \alpha R\approx \alpha \sqrt{A\over N_{CPU}}
\end{equation}
where $N_{CPU}$ is the number of CPUs. The number of time steps required to find a good load balance should typically be
\begin{equation}
{\Delta r\over\delta r}\approx {\sqrt{\mathrm{N_{cpu}}}\over\alpha}.
\end{equation}
For the 3D case it is
\begin{equation}
{\Delta r\over\delta r}\approx {\mathrm{N_{cpu}}^{1/3}\over\alpha}.
\end{equation}
The Voronoi based parallelization scheme is not limited to a particular implementation of solving the Euler equations, it can be applied to SPH, AMR and moving mesh codes. Extending the Voronoi based parallelization scheme is trivial for SPH, where each SPH particle is a hydro point in our notation. For AMR (or other grid based codes), each cell can be associated with a point (its location inside the cell is arbitrary but a natural choice would be the cells center of mass), thus enabling the Voronoi based parallelization scheme where the points that are inside the cells are the hydro points in our notation.
\section{Performance}
In the following section we run several benchmarks to test our method. All tests are using the RICH moving mesh hydrodynamics code (2014 In preparation), which solves the Euler equations on a moving Voronoi mesh. RICH is made parallel by using MPI and the Voronoi based parallelization scheme described in this paper. All of the tests are run on the Astric cluster in Hebrew University, which hosts 56 Intel E5-2670 Xeon processors for a combined total of 448 cores connected via InfiniBand FDR non-blocking connection. For all of the following tests the load balance is defined to be the ratio between the {MPI process with the highest number of cells to the average zones per MPI process, i.e. $\max{M_i}/M_{\mathrm{best}}$.
\subsection{Load Balancing}
We use a simple blast wave problem in order to check the performance of our new load balancing scheme.
The initial problem is set up with a random mesh of uniform density, with a total of $10^4\cdot128$ hydro points distributed equally among 128 MPI processes. In order to measure the effectiveness of our parallelization scheme, we plot the load balancing $\max{M_i}/ M_{\mathrm{best}}$ as a function of time in figure \ref{fig:load}. The test starts with almost perfect load balance, and as the hydro points begin to redistribute themselves, the work is no longer perfectly distributed. However, the average load balance is close to unity (mean value of 1.11), and thus the parallelization is efficient considering that there are no global redistributions of data.
\begin{figure}
  \plotone{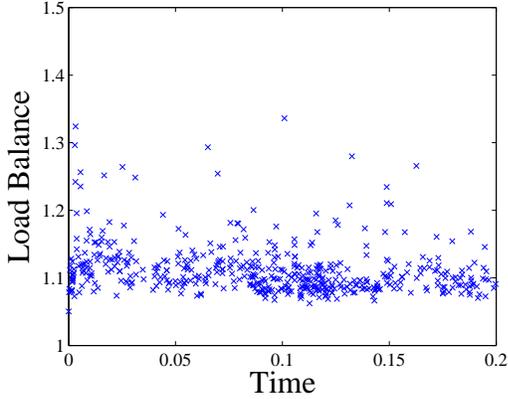}\\
  \caption{The load balance, defined as the ratio between the maximum number of hydro points in a MPI process to the average number of hydro points per MPI processes, as a function of time for the blast wave test problem. The simulation was run on 128 MPI processes with a total number of $10^4\cdot128$ hydro points.}\label{fig:load}
\end{figure}

\subsection{Weak Scaling}
In a perfect parallelization scheme, increasing the number of MPI processes while keeping the workload per MPI process constant should result with a constant computational time per time step; this is known as weak scaling. However, the need to communicate information between processes and load imbalance prevent this from happening in practice.

In order to check the weak scaling of our scheme, the same blast wave problem as in the previous section is run. First we keep a constant workload of $10^5$ hydro points per MPI process, while increasing the number of MPI processes, and a second time keeping $10^4$ hydro points per MPI processor. Figure \ref{fig:weak} shows that for a small number of MPI processes the time per time step increases slightly as the number of MPI processes is increased. There are two main factors for this. The first, is the ability of a core to increase its speed while the number of MPI processes that are being used on its CPU die is low (this is known as Turbo Boost). The second is that our implementation currently uses blocking messages when we send and receive data from the neighboring MPI processes. This gives rise to idle time when the MPI processor is waiting for a message. This inefficiency saturates once there are MPI processes that are completely surrounded by other MPI processes (as opposed to the domain walls). Future versions of RICH will use non-blocking messages, which should improve performance as well as other performance increasing changes. In order to show the effect of Intel's Turbo Boost, we run the single thread version of the code with 4-8 concurrent runs per CPU and measure their run time relative to a single run. We then rescale the results of the run with $10^5$ hydro points per MPI process accordingly and plot them in fig.\ref{fig:weak}. The ratio between our single thread with $10^5$ hydro points' run time and the rescaled result for 448 MPI process is 0.78. The ratio between the run time of 16 MPI processes and 448 MPI process is 0.93, showing that our code has good scaling once the number of neighbors saturates.
For comparison, we run the same blast wave problem with $10^5$ SPH points per MPI process with Gagdet2 \citep{Gadget}, using a fixed time step for all of the particles and in 3D (which is better optimized than the 2D version). The results are shown in fig. \ref{fig:weak}. Our code clearly scales better for more than 32 MPI processes, while our scaling for less than 32 MPI processes is worse.
\begin{figure}[h]
  \plotone{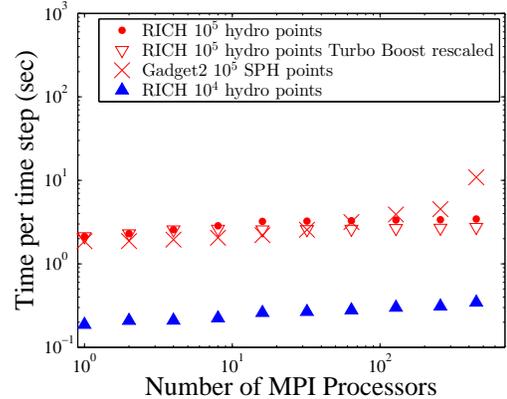}\\
  \caption{Weak scaling for the blast wave test problem. Plotted is the number of seconds per time step while keeping the workload per MPI process constant with $10^4$ hydro points, shown in blue triangles, and $10^5$ hydro points per MPI process, shown in red circles. The red upside triangles are the results for the $10^5$ hydro points but rescaled to take into account Intel's Turbo boost. For comparison, the results of Gadget2 are shown as red crosses.}\label{fig:weak}
\end{figure}

\subsection{Strong Scaling}
Ideally, for a fixed number of hydro points, the computational time should scale inversely proportional to the number of processors used. It should be equal to the computational
time on a single processor divided by the number of processors used. In practice this is not achieved due to imperfect load balancing and
time spent on communication between processors. Strong scaling, defined as keeping the total workload constant while the number of CPUs increases, is another way to measure the efficiency of the parallelization scheme.

The same problem as before is ran, once with a total workload of $10^6$ hydro points and once with $10^7$ hydro points. The same qualitative behavior as in the weak scaling can be seen in figure \ref{fig:strong}. For the larger workload, the scaling is one over the number of MPI processes. For the smaller workload, using more than 256 MPI processes decreases the efficiency. This is because the workload per MPI process is so low that the overhead in constructing the ghost cells needed for the boundaries (with their communication) dominates the computation. At least for this implementation, running with less than 4000 hydro points per MPI process is counterproductive.
\begin{figure}[h]
  \plotone{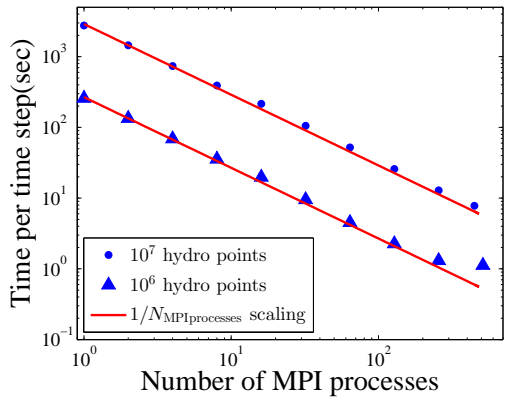}\\
  \caption{Strong scaling for our blast wave test problem. Plotted is the time per time step while keeping the global workload constant. Blue circles are used for the run with a global workload $10^7$ cells, blue triangles for a global workload of $10^6$ cells and the red lines show $1/N_{\mathrm{MPI_Processors}}$ scaling.}\label{fig:strong}
\end{figure}

\subsection{Uniform Distribution}
In this section we test the performance of this scheme on a domain which has uniform initial distribution of hydro points. We run the Gresho vortex problem \citep{Gresho} as set up in \cite{Arepo} on 64 MPI processes starting with a uniform distribution of cells while each MPI process hosts $10^4$ hydro points. Figure \ref{fig:load2} shows the load balancing, as defined above, as a function of time. It is evident that good efficiency is maintained at all times.
\begin{figure}[h]
  \plotone{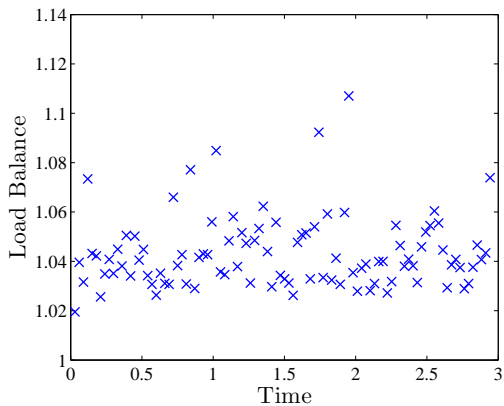}\\
  \caption{The load balance as a function of time for our Gresho vortex test problem.}\label{fig:load2}
\end{figure}

\subsection{Achieving Load Balance}
\label{sec:exp}
This scheme can also quickly achieve excellent load balance even when the initial distribution of work is very unbalanced.
To show this, we set up a static mesh problem, where  the hydro points are distributed in the domain $[-1,1]^2$, with random position picked from an exponential distribution that is described such that the probability of having a hydro point between $r$ and $r+dr$, where $r$ is the radius, is given by $\lambda\exp{(-\lambda r)dr}$ with $\lambda=10$.
The initial positions of CPU mesh points is randomly chose from a uniform distribution in the unit square.
This initial setup gives an inefficient configuration, and our goal is to show how the processors position move according to our algorithm and advance towards more efficient load balance. The hydro mesh points are kept fixed and, in fact, no hydro is calculated.

The system is then iterated using the load balancing steps described in the previous section, and the load balance is recorded for each iteration. This test is preformed with 96 MPI processes and a total of $9.6\cdot 10^5$ hydro mesh points.
With our initial setup the load balance is 66 and the configuration of the CPUs is shown in fig. \ref{fig:cpuload}. The load balance as a function of the iteration number is plotted in fig. \ref{fig:cpuiter} while the final configuration of the CPUs is shown in fig. \ref{fig:cpuload}.
Our scheme improves the load balance which eventually saturates on the average value of $1.4$.
It takes about $\sim 300\textrm{th}$ iterations to achieve this load balance, which is comparable to our estimate of $\sqrt{\mathrm{N_{cpu}}}/\alpha=100$. The mean value of the load balance after 300 iterations is 1.4, a dramatic improvement in efficiency over the initial load balance of 60.
\begin{figure}
\centering
% \begin{subfigure}[h]{0.48\textwidth}
%  \includegraphics[width=0.96\textwidth]{CPUbegin}
%  \end{subfigure}
%\begin{subfigure}[h]{0.48\textwidth}
%  \includegraphics[width=0.96\textwidth]{CPUend}
%   \end{subfigure}
\epsscale{2}
\plottwo{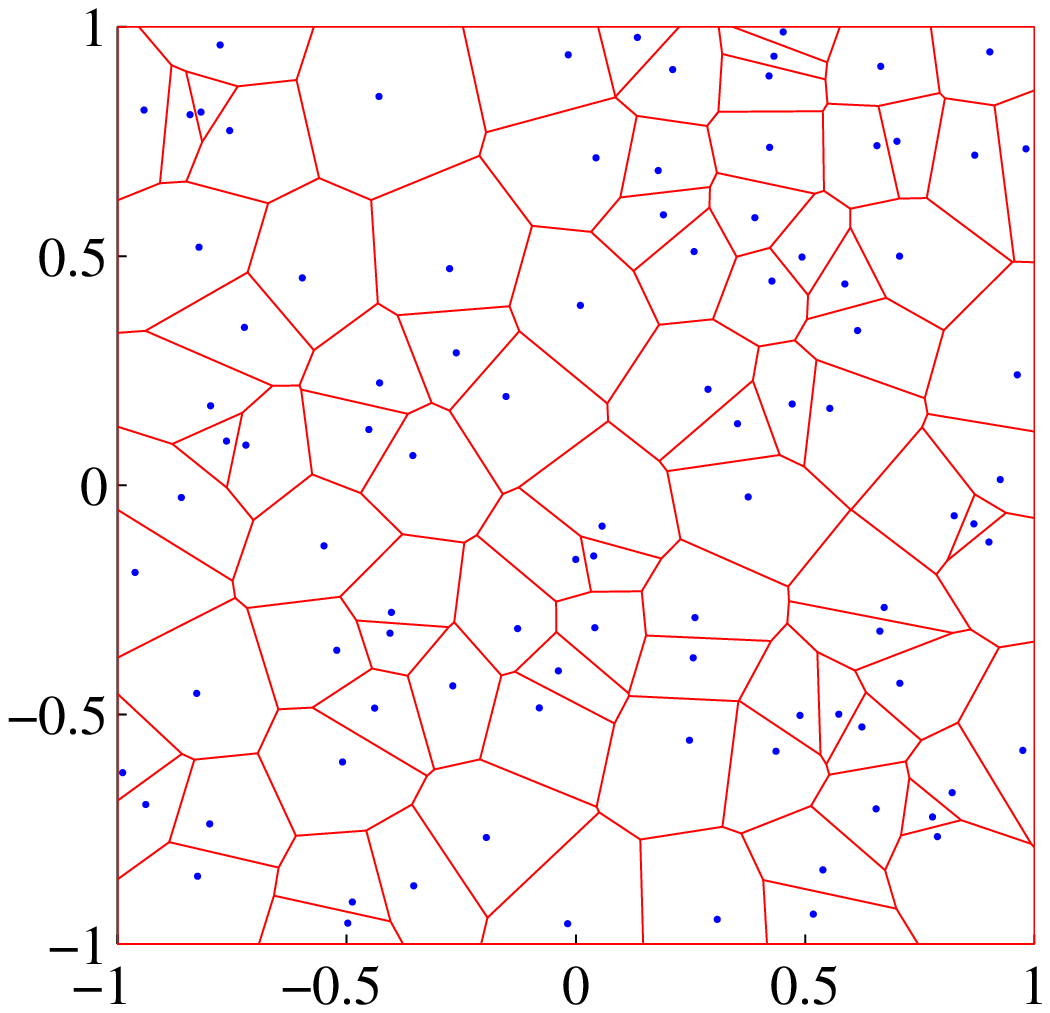}{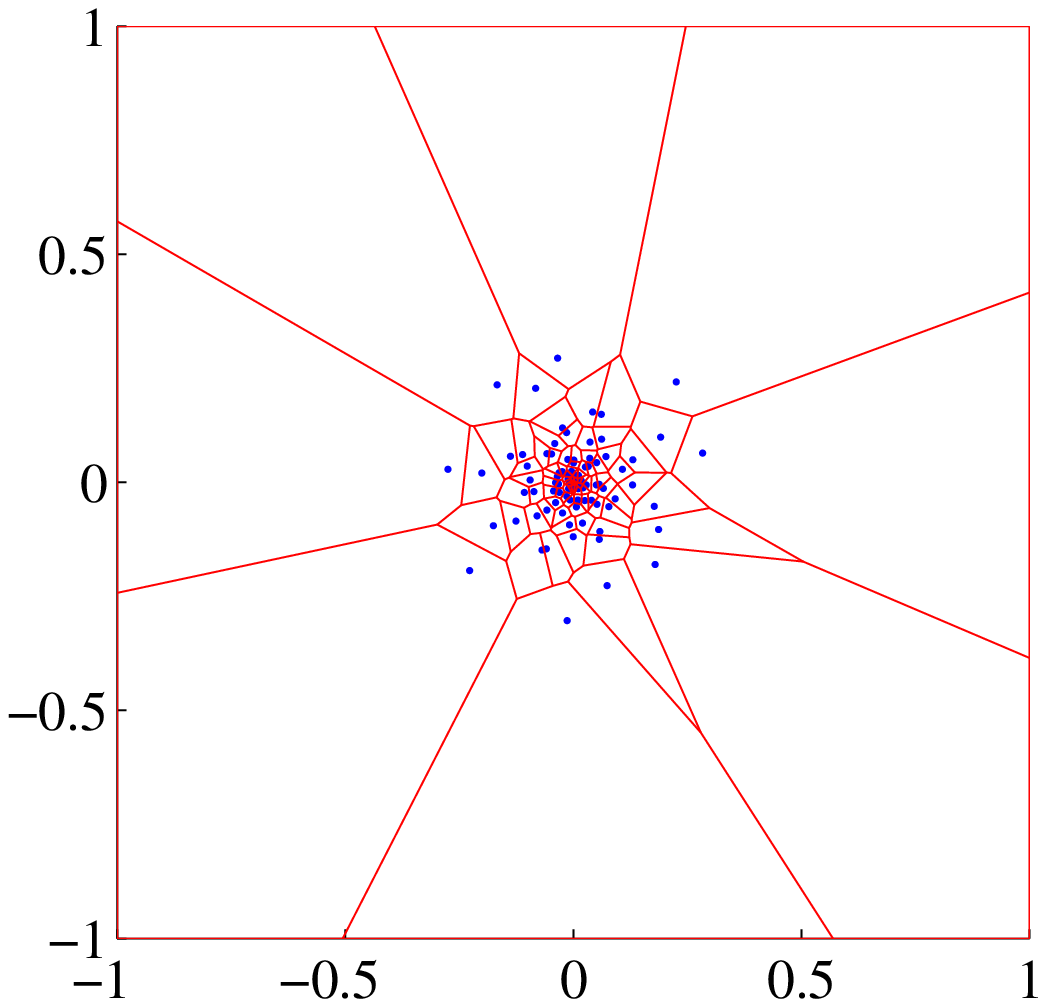}
   \caption{The initial (left) and final (right) setup of the CPUs for the exponential distribution load balance test.}
    \label{fig:cpuload}
\end{figure}
\begin{figure}[h]
  \plotone{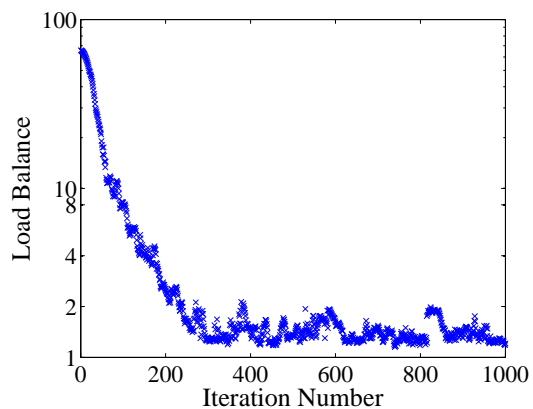}\\
  \caption{The load balance as a function of the iteration number for the exponential distribution load balance test.}\label{fig:cpuiter}
\end{figure}

\section{Further Extensions}
Our merit system can easily be extended to deal with more complex scenarios where the computational time per hydro point is not constant. Hydro points whose computational time is longer than other points can simply be taken into account as more than one hydro point when the effective pressure is calculated.
An effective implementation of this option is to calculate the time it took each CPU to advance a single time step and give a merit based on that time. This way, the algorithm will tend to even out the computational time among the CPUs even if additional calculations other than hydrodynamics are taking place.

In some cases a large dynamical range of time steps, which are different for different hydro points, is present \citep{Katz}. Some codes (e.g. \cite{Arepo}) have an option to advance in time only a subset of hydro points with their relevant time step, thus while one hydro point is advanced one large time step, another hydro point with a smaller time step can be advanced many smaller time steps. Defining a pressure just as the time it takes each CPU cell, will results in good load balance, but
could create a very large memory imbalance between CPUs, since some CPUs will have many hydro points with a large time step and some will have a small number of hydro points with small time steps. One way to overcome this issue is to use a MultiGrid approach. A Voronoi diagram of the processors should be created for every time step level, and then our method for load balancing should be applied to each time step level individually while taking into account only hydro points with the corresponding time step.
Note that there is no reason for the CPU to have its domains overlap between the different time step levels.

Since our load balancing method is a geometrical one, problems with extreme non-uniformity in the hydro points distribution can be difficult to address without allowing the Voronoi cells of the CPUs to have large aspect ratios. There are two possible improvements to handle these situations. The first is to have each CPU run more than one MPI process, this allows the CPU domains to be smaller at the cost of more communications. The second is to have an additional ``global'' attraction force that helps contract the CPU cells in areas of extreme hydro points density. As an example of a ``global" attraction force, the mesh points of the CPUs can be moved according to eq. 71 in \cite{Arepo}. Specifically we use
\begin{equation}
\vec{dx_i}=\sum_{j\neq i}\pi \left(\frac{R_i}{|\vec{x}_i^n-\vec{x}_j^n|} \right)^3\left(\frac{M_{\mathrm{best}}}{M_j}-1\right) \left(\vec{x}_i^n-\vec{x}_j^n\right)
\label{eq:global}
\end{equation}

As an example of the above, we run a problem similar to the one in section \ref{sec:exp}, but instead of having one center for the exponential  distribution, we have 40 different centers distributed in a uniform random manner, with $\lambda=320$. For each center, a total of $2\cdot 10^5$ hydro points are drawn. The initial locations of the CPUs are drawn uniformly. They are then evolved with our load balancing scheme with the addition of the above mentioned ``global" attractive force given by eq. \ref{eq:global}. We limit the magnitude of this force to be no larger than $\alpha R_i/5$. The load balance as a function of the iteration number is presented in fig. \ref{fig:galaxyload}. The mean load balance after 1000 iterations is 2.05, using 4096 MPI processes. This shows that our parallelization scheme can perform well on an extremely inhomogeneous problems with a large number of MPI processes.

\begin{figure}
\vspace{0.1 in}
 \plotone{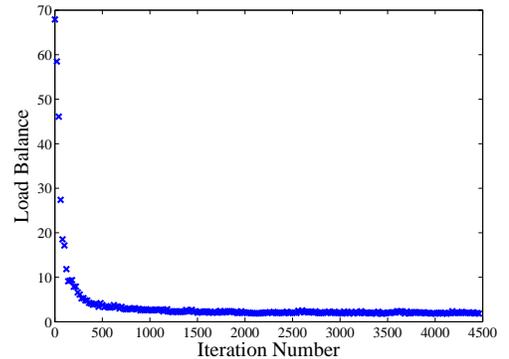}
  \caption{The load balance as a function of the iteration number for the exponential distribution load balance test.}
  \label{fig:galaxyload}
\end{figure}

\section{Discussion}
We present here a novel method to parallelize a computational code on a distributed memory machine. It is relevant for handling the load balance of partial differential equations that can be solved with domain decompositions. This method, which uses a moving Voronoi diagram to determine the CPU domains, ensures that there is no need for global data redistribution throughout the whole computation but only exchange of points adjacent to the edges of the CPU's domain. The ability of the CPUs to ``flow'' with the hydro points greatly reduces the need to transfer data between CPUs. Tests run with the RICH moving mesh code, show very good scaling up to 448 MPI processes.

The main caveat for this parallelization scheme is that it is efficient only as long as the number of CPUs is smaller than the number of hydro points per CPU. Since typically the number of hydro points per CPU is much larger than the number of CPUs, this caveat does not impose a strong constraint.
\begin{acknowledgments}
%\section*{Acknowledgement}
We would like to thank  Frederic Bournaud, Romain Teyssier and Jim Stone for their insightful comments. ES is supported by an Ilan Ramon grant from the
Israeli Ministry of Science. This research is supported in part by  ISF, ISA, iCORE grants and a Packard Fellowship.
\end{acknowledgments}
\bibliography{parallelpaper}
\end{document}